\documentclass[10pt,conference]{IEEEtran}
\newcommand{\CLASSINPUTinnersidemargin}{1in} 
\newif\ifpdf
\ifx\pdfoutput\undefined \else
  \ifx\pdfoutput\relax
  \else
    \ifcase\pdfoutput
    \else
      \pdftrue
    \fi
  \fi
\fi

\usepackage{cite}

\ifCLASSINFOpdf
   \usepackage[pdftex]{graphicx}
   \graphicspath{{.}}
\else
   \usepackage[dvips]{graphicx}
   \graphicspath{{.}}
\fi
\usepackage[cmex10]{amsmath}
\usepackage{algorithmic}
\usepackage{algorithm}

\usepackage{array}

\usepackage{mdwmath}
\usepackage{mdwtab}

\newtheorem{definition}{Definition}
\newtheorem{theorem}{Theorem}
\newtheorem{claim}{Claim}
\newtheorem{lemma}{Lemma}
\newtheorem{corollary}{Corollary}
\IEEEoverridecommandlockouts

\begin{document}
\title{Efficient and Robust Compressed Sensing using High-Quality Expander Graphs}

\author{\IEEEauthorblockN{Sina Jafarpour}
\IEEEauthorblockA{Computer Science\\
 Princeton University\\
sina@cs.princeton.edu}
\and
\IEEEauthorblockN{Weiyu Xu}
\IEEEauthorblockA{Electrical Engineering\\
 California Institute of Technology\\
weiyu@caltech.edu  
}
\and
\IEEEauthorblockN{Babak Hassibi \thanks{ "The work of B. Hassibi was supported in parts by the National Science Foundation under grant CCF 0729203, by the David and Lucille Packard Foundation and by Caltech's Lee Center for Advanced Networking."}
\IEEEauthorblockA{Electrical Engineering\\
California Institute of Technology \\
  hassibi@caltech.edu}}
\and
\IEEEauthorblockN{Robert Calderbank}
\IEEEauthorblockA{Electrical Engineering
\\Princeton University\\calderbk@math.princeton.edu}
}
\maketitle


\begin{abstract}
Expander graphs have been recently proposed to construct efficient
compressed sensing algorithms. In particular, it has been
shown that any $n$-dimensional vector that is $k$-sparse (with $k\ll
n$) can be fully recovered using $O(k\log\frac{n}{k})$ measurements
and only $O(k\log n)$ simple recovery iterations. In this paper we
improve upon this result by considering expander graphs with expansion
coefficient beyond $\frac{3}{4}$ and show that, with the same number
of measurements, only $O(k)$ recovery iterations are required, which
is a significant improvement when $n$ is large. In fact, full recovery can be
accomplished by at most $2k$ very simple iterations. The number
of iterations can be made arbitrarily close to $k$, and the recovery algorithm can be implemented very efficiently using
a simple binary search tree. We also show that by tolerating a
small penalty on the number of measurements, and not on the number of
recovery iterations, one can use the efficient construction of a
family of expander graphs to come up with explicit measurement matrices for this method. We compare our result with other
recently developed expander-graph-based methods 
and argue that it compares favorably both in terms of the
number of required measurements and in terms of the recovery time
complexity. Finally we will show how our analysis extends to give a robust algorithm  that finds the position and sign of the $k$ significant elements of  an almost $k$-sparse signal and then, using very simple optimization techniques, finds in sublinear time a $k$-sparse signal which approximates the original signal with very high precision.
\end{abstract}


\IEEEpeerreviewmaketitle

\section{Introduction}
The goal of \textit{compressive sampling} or \textit{compressed
  sensing} is to replace the conventional sampling and reconstruction
operations with a more general combination of linear measurement and optimization in order to acquire certain kinds of signals at a rate
significantly below  Nyquist.  Formally, suppose we have a
signal $x$ which is sparse. We can model $x$ as a $n$ dimensional
vector that has at most $k$ non-zero components . We desire to find an
$m\times n$ matrix $A$ such that m , the number of measurements, becomes as
small as possible ( and can be efficiently stored ) and x can be
recovered efficiently from $y=Ax$.

The originally approach was through the use of dense random
matrices and random projections. It has been shown that if the matrix
$A$ has some \textbf{restricted isometry property (RIP-2)}, that is, it almost
preserves the Euclidean norm of all $3k-sparse$ vectors, then $A$ can
be used in compressed sensing and the decoding can be accomplished using \textit{linear programming and convex programming} methods
\cite{candes}.  This is a geometric approach based on linear and
quadratic optimization, and \cite{JL} showed that the $RIP-2$
property is a direct consequence of the \textit{Johnson-Linderstrauss
  lemma} \cite{JL2} so that many dense random matrices will satisfy
this property. However, the problem in practice is that the linear and
quadratic programming algorithms have cubic complexity in $n$ and become really inefficient, as $n$ becomes very large; furthermore, in
order to store the whole matrix in memory we still need  $O(m\times
n)$ which is inefficient too.

Following \cite{XH1,XH2,berinde,newindyk,oldindyk}, we
will show how random dense matrices can be replaced 
by the adjacency matrix of a high quality family of expander graphs,
thereby reducing the space complexity of matrix storage and,
more important, the recovery time complexity to a few very
simple iterations. The main idea is that we study
expander graphs with expansion coefficient beyond the $\frac{3}{4}$
that was considered in \cite{XH1,XH2}. 

The remainder of the paper is organized as follows. In Section
\ref{sec:prev} we review the previous results from \cite{XH1,XH2}. In
Section \ref{sec:rip}, following the geometric approach of
\cite{berinde}, we establish that the adjacency matrix of the
expander graphs satisfies a certain Restricted Isometry Property for
Manhattan distance between sparse signals. Using this property,
or via a more direct alternative approach, we show how the recovery
task becomes much easier. In Section \ref{sec:res} we generalize the
algorithm of \cite{XH1,XH2} to expander graphs with expansion
coefficient beyond $\frac{3}{4}$. The key difference is that now the
progress in each iteration is proportional to $\log n$, as opposed to a
constant in \cite{XH1,XH2}, and so the time complexity is reduced from
$O(k\log n)$  to $O(k)$. We then describe how the algorithm can be
implemented using simple data structures very 
efficiently and show that explicit constructions
of the expander graphs impose only
a small penalty in terms of the number of measurements, and not the
number of iterations, the recovery algorithm requires. We also compare
our result to previous results based on 
random projections and to other approaches using the adjacency matrices
of expander graphs. In section \ref{sec:robust} generalize the analysis to a family of almost $k$-sparse signals; (after a few very simple iterations) the robust recovery algorithm proposed in \cite{XH1} empowered with high-quality expander graphs  finds the position and the sign of the $k$ significant elements of an almost $k$-sparse signal. Given this information, we then show how \textsl{the restricted isometry property} of the expander graphs lets us use very efficient sub-linear optimization methods to find a $k$-sparse signal that approximates the original signal very efficiently. Section
\ref{sec:conc} concludes the paper.


\section{Previous result: $O(k\log n)$ recovery}
\label{sec:prev}
\subsection{Basic Definitions}
Xu and Hassibi \cite{XH1} proposed a new scheme for compressed sensing with deterministic recovery 
guarantees based on combinatorial structures called \textit{unbalanced expander graphs}:
\begin{definition}[Bipartite Expander Graph, Informally]

 An expander graph \cite{salil2} $E$ is a $d$ regular graph with $v$ vertices, such that:
\begin{enumerate}
	\item $ E$ is sparse (ideally $d$ is much smaller than $v$).
	\item $E$ is \textit{" well connected"}. 
\end{enumerate}
\end{definition}
Various formal definitions of the second property define the various types of expander graphs. The expander graph used in  \cite{XH1}, \cite{XH2} which has suitable properties for compressed sensing is the "\textit{vertex expander}" or "\textit{unbalanced expander}" \textit{bipartite graph}:

\begin{definition}[Unbalanced Bipartite $\frac{3}{4}$-Expander Graph]
A bipartite left regular graph with $n$ variable nodes, $m$ parity
check nodes and left degree $d$ will be $(\alpha n, \frac{3}{4}d) $
expander graph, for $0<\alpha<\frac{1}{2}$, if for every subset of
variable nodes $\cal V$ with cardinality $|{\cal V}|\leq \alpha n$,
the number of neighbors connected to $\cal V$, denoted by $N({\cal
  V})$ is strictly larger than $\frac{3}{4}d {\cal V}$, i.e, $|N({\cal
  V})> \frac{3}{4} d |{\cal V}|$.  
\label{def:2}
\end{definition}

Using the probabilistic method, Pinsker and Bassylago \cite{BP} showed the existence of $\frac{3}{4}-$expander graphs and they showed that any random left-regular bipartite graph will, with very high probability, be an expander graph. Then Capalbo et al. gave an explicit construction for these expander graphs.

\begin{theorem}
\label{constexpand}
Let $0<r<1$ be a fixed constant. Then for large enough $n$ there exists a $(\alpha n, \frac{3}{4} d)$ expander graph $E$ with $n$ variable nodes and $\frac{n}{r}$ parity check nodes with constant left degree (not growing with $n$) $d$ and some $0<\alpha<1$. Furthermore, the explicit zig-zag construction can deterministically construct the expander graph.
\end{theorem}

Using Hoeffding's inequality and Chernoff bounds Xu and Hassibi \cite{XH2} showed the following theorem.
\begin{theorem}
\label{secondexpand}
For any $k$, if $n$ is large enough, there exists a left regular bipartite graph with left degree $d$ for some number $d$, which is  $(k, \frac{3}{4} d)$ expander graph with $m=O(k \log n)$ parity check nodes.
\end{theorem}
\subsection{Recovery Algorithm}
Suppose $\hat{x}$ is the original $n$ dimensional $k$-sparse signal,
and the adjacency matrix of a $(\alpha n, \frac{3}{4}d)$ expander
graph is used as the measurement matrix for the compressed sensing. We are given $y=A \hat{x}$ and we want to recover $\hat{x}$. Xu and Hassibi \cite{XH1} proposed the following algorithm:
\begin{algorithm}
 
  \caption{Left Degree Dependent Signal Recovery algorithm}
  \label{XH1}
   \begin{algorithmic}[1]
  \item [1)] Initialize $x=0_{n\times 1}$.
  \item [2)] IF $y=A x$ output $x$ and exit.
  \item [3)] ELSE find a variable node say $x_{j}$ such that more than half of the measurements it participate in, have identical gap $g$.
  \item [4)] set $x_{j} \leftarrow x_{j} + g $, and go to 2.
	\end{algorithmic}
\end{algorithm}

In the above algorithm the gap is defined as follows.

\begin{definition}[gap] Let $\hat{x}$ be the original signal and
  $y=A\hat{x}$. Furthermore, let $x$ be our estimate for
  $\hat{x}$. For each value $y_i$ we define a gap $g_i$ as: 
	$$g_i=y_i -\sum_{j=1}^n A_{ij}x_j.$$
\label{def:gap}
\end{definition}

Xu and Hassibi \cite{XH1} proved the following theorem that bounds the number of steps required by the algorithm to recover $\hat{x}$.
\begin{theorem}
\label{XHA1}
Suppose $A$ is the adjacency matrix of an expander graph satisfying Definition \ref{def:2}, and $\hat{x}$ is an $n$
dimensional $k$ sparse signal (with $k<\frac{\alpha n}{2}$), and
$y=A\hat{x}$. Then Algorithm 
\ref{XH1} will always find a signal $x$ which is $k$ sparse and for which
$Ax=y$. Furthermore, the algorithm requires at most $O(kd)$ iterations,
where $k$ is the sparsity level of the signal and $d$ is the left side
degree of the expander graph. 
\end{theorem}

Let us now consider the consequences of the above Theorem for the
expander graphs in Theorems 1 and 2. In Theorem 1 the sparsity can
grow proportional to $n$ (since $k<\frac{\alpha n}{2}$) and the
algorithm will be extremely fast; the algorithm requires
$O(kd)$ iterations and since $d$ is a constant independent of $n$, the number of iterations will be $O(k)$.  We also
clearly need $O(n)$ measurements. 

In Theorem 2 the sparsity level $k$ is fixed (does not grow with $n$)
and the number of measurements needs to be $O(k\log n)$, which is
desired. Once more the number of required iterations is
$O(kd)$. However, in this case Xu and Hassibi showed the following
negative result for $(k,\frac{3}{4}d)$ expander
graphs.
\begin{theorem}
\label{negative}
Consider a bipartite graph with $n$ variable nodes and $m$ measurement
nodes, and assume that the graph is a $(k,\frac{3}{4}d)$ expander
graph with regular left degree $d$. Then if $m=O(k \log n)$ we have $
d=\Omega(\log n)$. 
\end{theorem}

This theorem implies that for a $(k, \frac{3}{4}d)$
expander graph, the recovery algorithm needs $O(k \log n)$
iterations. The main contribution of the current paper is that the
number of iterations can be reduced to $O(k)$. The key idea is to use
expanders with expander coefficient beyond $\frac{3}{4}$. 

\textbf{Remark} Theorem \ref{XHA1} does not imply the
full  recovery of the sparse signal. It only states that the output of
the recovery 
algorithm will be a $k$ sparse signal $x$ such that $Ax=A\hat{x}$
where $\hat{x}$ is the original signal. However, in the next section we
show how an interesting property of the expander graphs called the $RIP-1$
property, implies the full recovery. We also give a direct proof by
showing that the null-space of the adjacency matrix of an expander
graph cannot be ``too sparse''. 

\section{Expander Codes, RIP-1 Property, and Full-Recovery Principle}
\label{sec:rip}
\subsection{Expander Codes}
Compressed Sensing has many properties in common with coding
theory. The recovery algorithm is similar to the decoding algorithms of
error correcting codes but over $R^m$ instead of a finite
field. As a result, several methods from coding theory have been
generalized to derive compressed sensing algorithms. Among these
methods are the generalization of Reed-Solomon codes by Tarokh
\cite{tarokh}, and very recent results by Calderbank, et al
\cite{ASSC}, which are based on second order Reed-Muller codes, and
Parvaresh et al \cite{PH}, based on list decoding.

In 1996, Sipser and Spielman \cite{SS} used expander graphs to build a family of linear error-correcting codes with linear
decoding time complexity. These codes belong to class of error
correcting codes called \textit{Low Density Parity Check (LDPC)
  Codes}. The work done by Xu and Hassibi is a generalization of these
expander codes to compressed sensing. Feldman et al \cite{Feldman} suggested a way of decoding expander
codes using linear programming, and linear programming is the usual recovery algorithm in compressed sensing. This leads to a better
understanding of compressed sensing using expander graphs and a very
different geometric perspective on the problem.  

\subsection{Norm one Restricted Isometry Property}
The standard Restricted Isometry Property, is an important condition that
enables compressed sensing using random projections. Intuitively, it says that the measurement almost preserves
the euclidean distance between any two sufficiently sparse
vectors. This property implies that recovery using $l_1$ minimization is
possible if a random projection is used for measurement. Indyk and
Berinde in \cite{berinde} showed that expander graphs satisfy a very
similar property called \textit{RIP-1} which states that if the
adjacency matrix of an expander graph is used for measurement, then
the Manhattan ($l_1$) distance between two sufficiently sparse
signals will be preserved by measurement. They used this property to
prove that $l_1$-minimization is still possible in this case. However, we
will show in this section how RIP-1 can guarantee that the 
algorithm described will have full recovery. 

Following \cite{berinde}, we will show that the RIP-1 property can be
 derived from the expansion 
 property and will guarantee that if $\hat{x}$ is the original
 $k$-sparse signal, then no recovery algorithm can output a $k$-sparse
 signal $x$ such that $x \neq \hat{x}$ but $Ax=A\hat{x}$.  
 
We begin with the definition of expander graphs with expansion
coefficient $1-\epsilon$, bearing in mind that we will be interested in
$1-\epsilon >\frac{3}{4}$. 

\begin{definition}[Unbalanced Expander Graph]
\label{genexpand}
 A $(l,1-\epsilon)$-\textit{unbalanced bipartite expander graph} is a
 bipartite graph $ E=(A,B), |A|=$ $n , |B|=$ $m$, where $A$ is the set
 of variable nodes and $B$ is the set of parity nodes, with regular
 left degree $d$ such that for any $X \subset A$, if $|X|\leq l$ then
 the set of neighbors $N(X)$ of $X$ has size $N(X)>(1-\epsilon)d|X|$.  
\end{definition}
 
The following claim can be derived using the Chernoff bounds\cite{berinde}\footnote{This claim is also used in the expander codes construction} :
\begin{figure}[!t]
\centering
\includegraphics[width=1.8in]{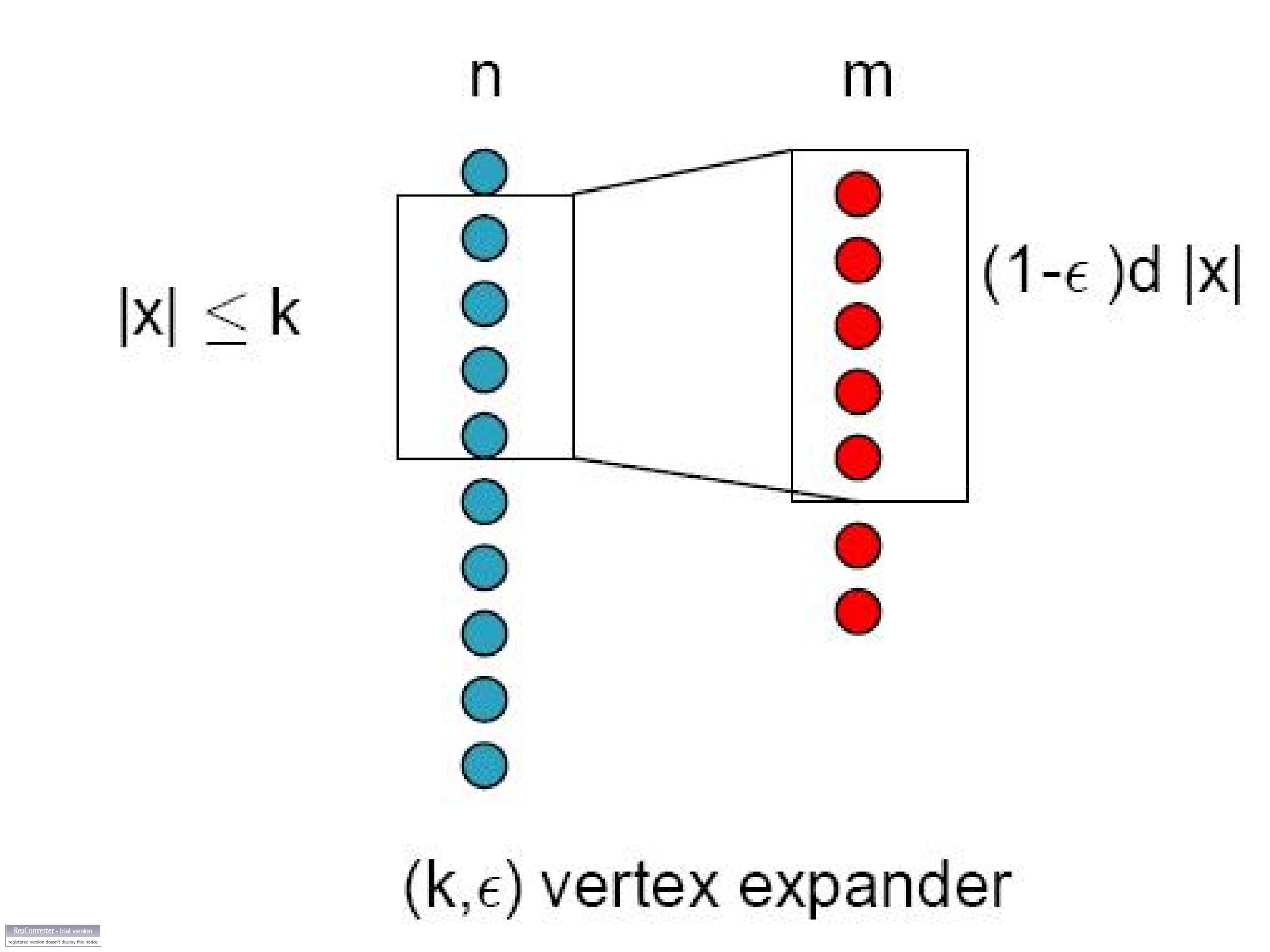}
\caption{$(k,\epsilon)$ vertex expander graph}
\label{figexp}
\end{figure}

\begin{claim}
for any $\frac{n}{2}\geq l \geq 1$ , $\epsilon>0$ there exists a
$(l,1-\epsilon)$ expander with left degree:
$$d=O\left(\frac{\log(\frac{n}{l})}{\epsilon}\right)$$ and right set
size: $$m=O\left(\frac{l\log(\frac{n}{l})}{\epsilon^2}\right)$$. 
\end{claim}

\begin{lemma}[RIP-1 property of the expander graphs]
\label{RIP}
 Let $A_{m \times n}$ be the adjacency matrix of a $(k,\epsilon)$ expander graph $E$, then for any $k$-sparse vector $x\in {\cal R}^n$ we have:
\begin{equation}
(1-2\epsilon)~ d ~||x||_1  \leq ||Ax||_1 \leq d~ ||x||_1
\end{equation}
\begin{IEEEproof}
The upper bound is trivial using the triangle inequality, so we only prove the lower bound:

The left side inequality is not influenced by changing the position of the coordinates of $x$, so we can assume that they are in a non-increasing order: $|x_1|\geq |x_2| \geq ... \geq |x_n|$. Let $(x_i,y_j)$ be the edge $e$ that connects $x_i$ to $y_j$. Define $E_2=\{(x_i,y_j):\exists k<i: (x_k,y_j)\in E\}$. Intuitively $E_2$ is the set of the collision edges. Let $T_i=\{e:\exists j~ s.t:~ e=(x_i,y_j), e \in E_2\}$ and $a_i=|T_i|$. \\Clearly $a_1=0$; moreover by the expansion property of the graph we have: for all  $k'\leq k: a_{k'}\leq \epsilon dk'$, and finally since the graph is $k$-sparse we know that for all $k''>k:$ $~x_{k''}=0$. Hence
\begin{IEEEeqnarray}{rcl}
	\sum_{(x_i,y_j)\in E_2} |x_i| &=& {\sum_{i=1}}^n |x_i| (a_i-a_{i-1})\IEEEnonumber\\&=& \sum_{i\leq k} a_i (|x_i|-|x_{i+1}|)\IEEEnonumber\\ &\leq& \sum_{i\leq k}	\epsilon di (|x_i|-|x_{i+1}|)\IEEEnonumber \\ &\leq& \sum_{i\leq k} |x_i| \epsilon d \IEEEnonumber\\ & =& \epsilon d \left|x\right|_1. \IEEEnonumber
\end{IEEEeqnarray} 
Now the triangle inequality implies:
\begin{IEEEeqnarray}{rcl}
\left|Ax\right|_1 &\geq & |\sum_{ e \in E} x_i|\IEEEnonumber\\  &\geq& \sum_{ e \notin E_2} |x_i| +  \sum_{ e \in E_2} |x_i | - 2 \sum_{e \in E_2} | x_i | \IEEEnonumber\\ &\geq& d \left|x\right|_1 - 2\epsilon d \left|x\right|_1. \IEEEnonumber \\ &=& (1-2\epsilon)~d~||x||_1 \IEEEnonumber.
\end{IEEEeqnarray} 
\end{IEEEproof}
\end{lemma}

\subsection{Full Recovery}
The full recovery property now follows immediately from Lemma \ref{RIP}.
\begin{theorem}[Full recovery]
\label{full}
Suppose $A_{m \times n}$ is the adjacency matrix of a $(3k,1-\epsilon)$ expander graph. And $x_1$ is a $k-sparse$ and $x_2$ is a   $2k$-sparse vector, such that $Ax_1=Ax_2$ then $x_1=x_2$.
\begin{IEEEproof}
Let $z=x_1-x_2$. Since $x_1$ is $k-sparse$ and $x_2$ is $2k$-sparse, $z$ is $3k$-sparse\footnote{$||z||_0 \leq ||x_1||_0 + ||x_2||_0 = 3k$}. By Lemma \ref{RIP} we have:
$$||x_1-x_2||_1 \leq \frac{1}{(1-2\epsilon)~d} ||Ax_1-Ax_2||_1 = 0,$$
hence $x_1=x_2$.
\end{IEEEproof}
\end{theorem}

Note that the proof of the above theorem essentially says that the
adjacency matrix of a $(3k,1-\epsilon)$ expander graph does not have a
null vector that is $3k$ sparse. We will also give a direct proof of
this result (which does not appeal to RIP-1) since it gives a flavor
of the arguments to come.

\begin{lemma}[Null space of $A$] Suppose $A_{m \times n}$ is the
  adjacency matrix of a $(3k,1-\epsilon)$ expander graph with
  $\epsilon\leq\frac{1}{2}$. Then any
  nonzero vector in the null space of $A$, i.e., any $z\neq 0$ such
  that $Az = 0$, has more than $3k$ nonzero entries.
\end{lemma}
\begin{IEEEproof}
Define ${\cal S}$ to be the support set of $z$. Suppose that $z$ has at most $3k$ nonzero entries, i.e., that $|{\cal S}|\leq
3k$. Then from the expansion property we have that $N({\cal S}) >
(1-\epsilon)d|{\cal S}|$. Partitioning the set $N({\cal S})$ into the
two disjoint sets $N_1({\cal S})$, consisting of those nodes in
$N({\cal S})$ that are connected to a single node in ${\cal S}$, and
$N_{>1}({\cal S})$, consisting of those nodes in $N({\cal S})$ that
are connected to more than a single node in ${\cal S}$, we may write
$N_1({\cal S})+N_{>1}({\cal S})>(1-\epsilon)d|{\cal S}|$. Furthermore,
counting the edges connecting ${\cal S}$ and $N({\cal S})$, we have
$|N_1({\cal S})|+2|N_{>1}({\cal S})|\leq d|{\cal S}|$. Combining these
latter two inequalities yields $|N_1({\cal S})|>(1-2\epsilon)d|{\cal
  S}|\geq 0$. This implies that there is at least one nonzero element in $z$
that participates in only one equation of $y = Az$. However, this
contradicts the fact that $Az=0$ and so $z$ must have more than $3k$
nonzero entries.
\end{IEEEproof}

\section{Our results: Efficient Full Recovery}
\label{sec:res}
\subsection{Efficient $O(k \log n)$ measurement with $O(k)$ iteration recovery} 
In this section, we show the general unbalanced bipartite expander
graphs introduced in Definition \ref{genexpand} work much better than
 $\frac{3}{4}$-expanders, in the sense that it gives
the measurement size $O(k \log n)$ which is up to a constant the optimum
measurement size, and simultaneously yields a recovery algorithm which
needs only $O(k)$ simple iterations. 

Before proving the result, we introduce some notations used in the
recovery algorithm and in the proof. 

\begin{definition}[gap] Recall the definition of the gap from
  Definition \ref{def:gap}. At each iteration $t$, let $G_t$ be the
  support\footnote{set of nonzero elements} of the gaps vector at
  iteration $t$ : $$G_t=\mbox{support }(\vec{g_t})=\{i|y_i\neq
  \sum_{j=1}^n A_{ij}x_j\}.$$	
\end{definition}

\begin{definition}
At each iteration t, we define $S_t$ an indicator of the difference between the estimate $\hat{x}$ and $x$  : $$S_t=\mbox{support }(\hat{x}-x)=\{j:\hat{x_j}\neq x_j\}.$$ 
\end{definition}
Now we are ready to state the main result:
\begin{theorem}[Efficient and Certain Compressive Sampling ]
\label{recovery}
Let $\epsilon < \frac{1}{4}$ and suppose $A_{m \times n}$, as defined in definition \ref{genexpand} where $m=O\left(\frac{k \log n}{\epsilon^2}\right)$, be the adjacency matrix of a $(3k,\epsilon)$ expander graph. If we use $A$ as the measurement matrix in compressed sensing of $k$-sparse signals, the following algorithm \ref{ouralg} will recover the original signal $k$ sparse signal $x$ from its measured sketch $y=Ax$ with certainty using at most $O(k)$ simple iterations . 
\begin{algorithm}
   \caption{Our $O(k)$ iteration, recovery algorithm}
  \label{ouralg}
   \begin{algorithmic}[1]
  \item [1)] Initialize $x=0_{n\times 1}$.
  \item [2)] IF $y=A x$ output $x$ and exit.
  \item [3)] ELSE find a variable node say $x_{j}$ such that at least $(1-2\epsilon)~d$ of the measurements it participate in, have identical gap $g$.
  \item [4)] set $x_{j} \leftarrow x_{j} + g $, and go to 2.
	\end{algorithmic}
\end{algorithm}
\end{theorem}
The proof is virtually identical to that of \cite{XH1}, except that
we consider a general $1-\epsilon$ expander, rather than a
$\frac{3}{4}$-expander, and consists of the following lemmas. 
\begin{itemize}
	\item The algorithm never gets stuck, and one can always find a coordinate $j$ such that $x_j$ is connected to at least $(1-2\epsilon)d$ parity nodes with identical gaps.
	\item With certainty the algorithm will stop after at most $2k$ rounds. Furthermore, by choosing  $\epsilon$ small enough the number of iterations can become arbitrarily close to $k$. 
	\end{itemize}

\begin{lemma}[progress] 
\label{progress}
Suppose at each iteration $t$, $S_t=\{j:\hat{x_j}\neq x_j\}$. If $|S_t| < 2k$ then always there exists a variable node $x_j$ such that at least $(1-2\epsilon)d$ of its neighbor check nodes have the same gap $g$.
\begin{IEEEproof}
we will prove that there exists a coordinate $j$, such that $x_j$ is connected to at least $(1-2\epsilon)d$ check nodes uniquely, in other words no other variable node is connected to these nodes. This immediately implies the lemma.
\\Since $|S_t| < 2k$ by the expansion property of the graph $N(S_t)\geq (1-\epsilon)d |S_t|$. Now we are going to count the neighbors of $S_t$  in two ways. Figure \ref{exp2} shows the progress lemma.  \\We partition the set $N(S_t)$ into two disjoint sets: 
\begin{figure}[!t]
\centering
\includegraphics[width=1.3in]{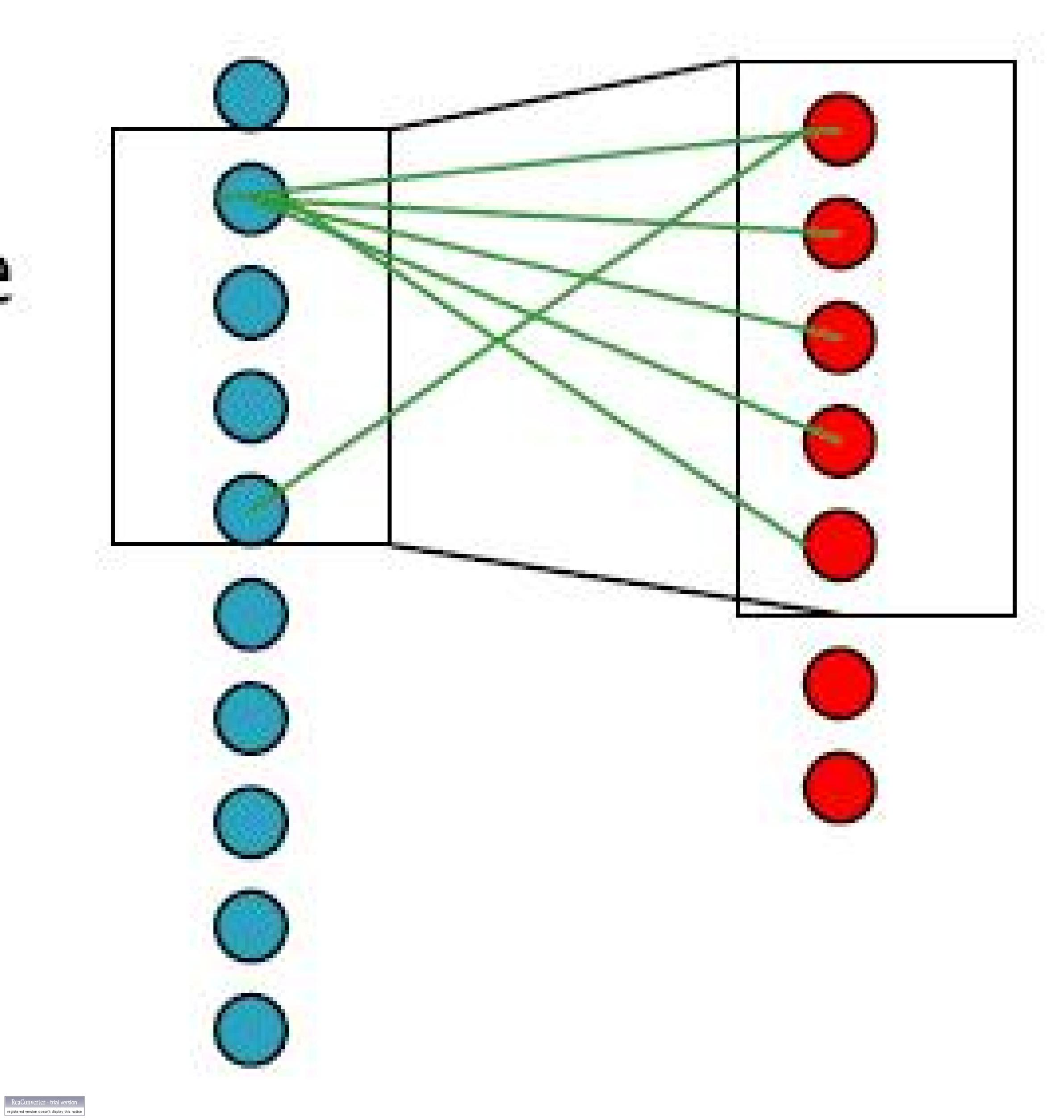}
\caption{Progress Lemma}
\label{exp2}
\end{figure}

\begin{itemize}
	\item $N_1(S_t)$: The vertices in $N(S_t)$ that are connected only to one vertex in $S_t$.
	\item $N_{>1}(S_t)$: The other vertices (that are connected to more than one vertex in $S_t$).
\end{itemize}
By double counting the number of edges between variable nodes and check nodes we have:
\begin{equation}
	|N_1(S_t)|+|N_{>1}(S_t)| =| N(S_t)| > (1-\epsilon) d |S_t| \IEEEnonumber
\end{equation}
\begin{equation}
	|N_1(S_t)|+2 |N_{>1}(S_t)| \leq \mbox{\#edges between } S_t,N(S_t) = d |S_t| \IEEEnonumber
\end{equation}
This gives
\begin{equation}
	 |N_{>1}(S_t)|  < \epsilon d |S_t|, \IEEEnonumber
\end{equation}
hence
\begin{equation}
	 |N_1(S_t)| > (1-2\epsilon) d |S_t|, \IEEEnonumber
\end{equation}
so by the pigeonhole principle, at least one of the variable nodes in $S_t$ must be connected uniquely to at least $(1-2\epsilon)d$ check nodes.
\end{IEEEproof}
\end{lemma}
\begin{lemma}[gap elimination]
\label{gapel}
At each step $t$ if $|S_t|< 2k$ then $|G_{t+1}| < |G_t|-(1-4\epsilon)d$
\begin{IEEEproof}
By the previous lemma, if $|S_t|<2k$, there always exists a node $x_j$
that is connected to at least $(1-2\epsilon)d$ nodes with identical
nonzero gap , and hence to at most $2\epsilon d$ nodes possibly with
zero gaps. Setting the value of this variable node to zero, sets the
gaps on these uniquely connected neighbors of $x_j$ to zero, but it
may make some zero gaps on the remaining $2\epsilon d$ neighbors
non-zero. So at least $(1-2\epsilon)d$ coordinates of $G_t$ will
become zero, and at most $2\epsilon d$ its zero coordinates  may
become non-zero. Hence \begin{equation}|G_{t+1}| < |G_t| -
  (1-2\epsilon)d +2\epsilon d = |G_t|-(1-4\epsilon)d. \end{equation} 
\end{IEEEproof}
\end{lemma}

{\bf Remark:} The key to accelerating the algorithm is the above
Lemma. For a $\frac{3}{4}$ expander, $\epsilon  = \frac{1}{4}$ and so
$|G_{t+1}| < |G_t|$, which only guarantees that $|G_{t+1}|$ is reduced
by a constant number. However, when $\epsilon  < \frac{1}{4}$, we have
$|G_{t+1}| < |G_t|-(1-4\epsilon)d$, which means that $|G_{t+1}|$ is
guaranteed to decrease proportionally to $d$. Since $d = \Omega(\log
n)$, we save a factor of $\log n$.

\begin{lemma}[preservation]
\label{preservation}
At each step $t$ if $|S_t|< 2k$, after running the algorithm we have $|S_{t+1}|< 2k$.
\begin{IEEEproof}
Since at each step we are only changing one coordinate of $x$, we have $|S_{t+1}|=|S_t|+1$, so we only need to prove that $S_{t+1}\neq 2k$.

Suppose for a contradiction that $|S_{t+1}|=2k$, and partition $N(S_{t+1})$ into two disjoint sets:
\begin{enumerate}
	\item $N_1(S_{t+1})$: The vertices in $N(S_{t+1})$ that are connected only to one vertex in $S_{t+1}$.
	\item $N_{>1}(S_{t+1})$: The other vertices (that are connected to more than one vertex in $S_{t+1}$).
\end{enumerate}
The argument is similar to that given above; by double counting the number of vertices in  $N_1(S_{t+1}),N_{>1}(S_{t+1})$ one can show that 	
\begin{equation}
|N_1(S_{t+1})|\geq (1-2\epsilon)~d ~2k \IEEEnonumber
\end{equation}
Now we have the following facts:
\begin{itemize}
	\item $|N_1(S_{t+1})| \leq |G_{t+1}|$ : Coordinates in $N_1(S_{t+1})$ are connected uniquely to coordinates in $S_{t+1}$, hence each coordinate in $N_1(S_{t+1})$ has non-zero gap.
	\item $|G_{t+1}| < |G_{1}|$: gap elimination from Lemma \ref{gapel}.
	\item $ |G_{1}| \leq kd $: $x$, $\hat{x}$ differ in at most $k$ coordinates, so $Ax, A\hat{x}$ can differ in at most $kd$ coordinates.
\end{itemize}
As a result we have:
\begin{eqnarray}
(1-2\epsilon)2~dk \leq |N_1(S_{t+1})| \leq |G_{t+1}| < |G_{1}| \leq kd 
\end{eqnarray}
This implies $\epsilon \geq \frac{1}{4}$ which contradicts the assumption  $\epsilon < \frac{1}{4}$.
\end{IEEEproof}
\end{lemma}

\begin{IEEEproof}[Proof of the Theorem \ref{recovery}]
Preservation  (Lemma \ref{preservation}) and progress (Lemma \ref{progress}) together immediately imply that the algorithm will never get stuck. Also by Lemma \ref{gapel} we had shown that $|G_1|\leq kd$ and $|G_{t+1}|< |G_t|-(1-4\epsilon)d$. Hence after at most $T=\frac{k}{1-4\epsilon}$ steps we will have $|G_T|=0$ and this together with the preservation lemma implies that we have discovered a signal $x$ such that $x$ is $2k$-sparse and $Ax=y$. Now since we had used a $(3k, \epsilon)$ expander, the full recovery property (Theorem \ref{full}) guarantees the recovery of the original signal.

Note that we have to choose $\epsilon < \frac{1}{4}$, and as an example, by setting $\epsilon=\frac{1}{8}$ the recovery needs at most $2k$ iterations.
\end{IEEEproof}
\textbf{Remark:} The condition $\epsilon < \frac{1}{4}$ in the theorem is necessary. Even $\epsilon=\frac{1}{4}$ leads to a $\frac{3}{4}$ expander graph (Definition \ref{secondexpand}), which needs $O(k\log n)$ iterations. 
\subsection{Explicit Construction of Expander Graphs}
In the definition of the expander graphs (Definition \ref{genexpand}), we noted that probabilistic methods prove that such expander graphs exist and furthermore, that any random graph will, with high probability, be an expander graph. Hence, in practice it may be sufficient to use random graphs instead of expander graphs. 

Though, there is no efficient explicit construction for the expander graphs of Definition \ref{genexpand}, there exists an explicit construction for a class of expander graphs which are very close to the optimum expanders of Definition \ref{genexpand}. Recently \cite{PV2}, Guruswami et al  based on Parvaresh-Vardy codes \cite{PV}, proved the following theorem:
\begin{theorem}[Explicit Construction of expander graphs]
\label{explicit}
For any constant $\alpha>0$, and any $n,k, \epsilon>0$, there exists a
$(k,1-\epsilon)$ expander graph with left degree:
$$d=O\left({\left(\frac{\log n}{\epsilon}\right
  )}^{1+\frac{1}{\alpha}}\right)$$ and number of right side vertices:
$$m=O(d^2 k^{1+\alpha})$$ which has an efficient deterministic
explicit construction. 
\end{theorem}
Since our previous analysis was only based on the expansion property, which does not change in this case, a similar result holds if we use these expanders. For instance by letting $\alpha=1$ and $\epsilon=\frac{1}{8}$ we will have an explicit expander construction with $d=O\left((\log n)^2\right)$ and so we just need  $m=O(k^2d^2 )$ number of measurements, and at most $2k$ number of iterations in the recovery algorithm.
\subsection{Comparison with the recent unified geometric-combinatorial approach}
We will compare our result with a very recent result by Indyk et al \cite{newindyk}. Their result unifies Indyk's previous work which was based on randomness extractors \cite{oldindyk} and a combinatorial algorithm with another approach to the RIP-1 property of Indyk-Berinde \cite{berinde} which is based on geometric convex optimization methods and suggests a recursive recovery algorithm which takes $m'=O(m \log n)=O(k \log^2 n)$ sketch measurements and needs a recovery time $O(m \log^2 n)=O(k \log^3 n)$.

By comparison, our recovery algorithm is a simple iterative algorithm, that needs $O(k \log n)$ sketch measurements. Our decoding algorithm consists of at most $2k$ very simple iterations. Each iteration can be implemented very efficiently (see \cite{XH1} ) since the adjacency matrix of the expander graph is sparse with all entries 0 or 1. One naive way to do that is by using  balanced binary search trees \footnote{such as red-black trees}.  As shown before, initially $|G_1|\leq kd$ so we can build the tree efficiently in $O(kd \log d)$. Now by gap elimination (Lemma \ref{gapel}), although at each iteration some nodes are going to be deleted from the tree and some new nods are added, the size of the tree does not grow, so all the updates can be done in $O(d \log d)$ . As a result, we have a $O(kd \log d)$ preprocessing step, and $2k$ iterations each taking $O(d \log d)$. So this naive approach has overall time complexity $O(kd \log  d)= O(k \log n \log\log n)$. This can even be improved by using better data structures.

\section{Approximately sparse signals and robust recovery}
\label{sec:robust}
In this section we will show how the analysis using high-quality expander graphs that we proposed in the previous section can be used to show that the robust recovery algorithm in \cite{XH1} can be done more efficiently in terms of the sketch size and recovery time for a family of almost $k$-sparse signals. With this analysis we will show that the algorithm will only need $O(k \log n)$ measurements. Explicit constructions for the sketch matrix exist and the recovery consists of two simple sub-linear steps. First, the combinatorial iterative algorithm in \cite{XH1} , which is now empowered with the high-quality expander sketches, can be used to find the position and the sign of the $k$ largest elements of the signal $x$. Using an analysis similar to the analysis in section \ref{sec:res} we will show that the algorithm needs only $O(k)$ iterations, and similar to the previous section, each iteration can be done efficiently using a variation of red-black trees and will have time complexity $O(d \log d)=O(\log n \log\log n)$.  Then restricting to the position of the $k$ largest elements, we will use a robust theorem in expander graphs to show that simple optimization methods that are now restricted on $k$ dimensional vectors can be used to recover a $k$ sparse signal that approximates the original signal with very high precision. In summary, both the combinatorial part and the optimization part require sub linear time complexity so the overall algorithm needs sub linear recovery time and will output a $k$-sparse signal very close to the original signal.

Before presenting the algorithm we will define precisely precisely what we mean for a signal to be almost k sparse.

\begin{definition}[almost $k$-sparse signal]
A signal $x\in R^n$ is said to be almost $k$-sparse iff it has at most $k$ large elements and the remaining elements are very close  to zero and have very low magnitude. In other words, the entries of the 'near-zero' level in the signal vector are near-zero elements taking values from the
set $[-\lambda,\lambda]$ while the 'significant' level of entries take
values from the set $S=\{x: |L-\Delta|\leq |x| \leq |L+\Delta \}$. By the definition of the almost sparsity we have $|S|\leq k$. The general assumption for almost sparsity is intuitively the fact that the total magnitude of the almost sparse terms should be small enough that so that it does not disturb  the overall structure of the signal which may make the  recovery  impossible or very errornous. Since $\sum_{x\notin S} |x| \leq n\lambda$ and the total contribution of the 'near-zero' elements is small we can assume that $n\lambda$ is small enough. We will use this assumption throughout this chapter.
\end{definition}

In order to make the analysis for almost $k$-sparse signals simpler  we will use a high quality expander graph which is right-regular as well\footnote{the right-regularity assumption is just for the simplicity of the analysis and as we will discuss it is not mandatory.}. The following lemma which is proved in \cite{raz} gives us a way to construct right-regular expanders from any expander graph without disturbing its characteristics (lemma 2.3 in \cite{raz}.

\begin{lemma}[right-regular expanders]
From any left-regular $(k,1-\epsilon)$ unbalanced expander graph $G$ with left size $n$, right size $m$, and left degree $d$ it is possible to efficiently construct a left-right-regular $(k,1-\epsilon)$ unbalanced expander graph $H$ with left size $n$, right size $m'\leq 2m$, left side degree $d' \leq 2d$, and right side degree $D=[\frac{nd}{m}]$ 
\end{lemma}
\begin{corollary}
There exists a $(k,1-\epsilon)$ left-right unbalanced expander graph with left side size $n$, right side size $m=O(k \log n)$, left side degree $d=O(\log n)$, right side degree $D=O(\frac{n \log n}{k \log n})=O(\frac{n}{k})$. Also based on the explicit constructions of expander graphs, explicit construction for right-regular expander graphs exists as well.
\end{corollary}

We will use the above right-regular high-quality expander graphs in order to perform robust signal recovery efficiently. The following algorithm generalizes the $k-sparse$ recovery algorithm and can be used to find the position and sign of the $k$ largest elements of an almost $k$-sparse signal  $x$ from $y=Ax$. Throughout the algorithm at each iteration $t$ let $\rho_t = 2t \Delta + (D-t-1) \lambda $ and $\phi_t=2t \Delta +(D-t) \lambda $. where $D=O(n)$ is the right side degree of the expander graph. Throughout the algorithm we will assume that $L> 2k \Delta +D \lambda$. Hence the algorithm is appropriate for a family of almost $k$-sparse signals for which  the magnitude of the significant elements is large enough. We will assume that $k$ is a small constant; when $k$ is large with respect to $n$, ($k=\theta(n)$),  the $(\alpha n, \frac{3}{4})$ constant degree expander sketch proposed in \cite{XH1} works pretty well.

\begin{algorithm}
   \caption{ The $O(k)$ iteration, robust recovery algorithm to find the position and sign of the $k$ largest elements of an almost-$k$-sparse signal $x$ and then a close $k$-sparse approximation for it. }
  \label{ouralgrob}
   \begin{algorithmic}[1]
  \item [1)] Initialize $\hat{x}=0_{n\times 1}$.
  \item [2)] At each iteration t, if $|y-A \hat{x}|_{\infty} \leq \phi_t$ determine the positions
and signs of the significant components in $x$ as
the positions and signs of the non-zero signal
components in $\hat{x}$; go  to 5.
  \item [3)] ELSE find a variable node say $\hat{x_{j}}$ such that at least $(1-2\epsilon)~d$ of the measurements it participate in are in either of the following categories:
 \begin{enumerate} 
  \item[a)] They have gaps which are of the same sign and have absolute values between $L-\Delta - \lambda - \rho_t$ and $L + \Delta + \lambda + \rho_t$.
 Moreover, there exists a number $G\in \{0, L+\Delta, L-\Delta\}$ such that $|y-A.\hat{x}|$ are all $\leq \phi_t$ over these $(1-2\epsilon)~d$ measurements if we change $\hat{x_j}$ to $G$.
 \item[b)] They have gaps which are of the same sign and have absolute values between $2L-2\Delta - \rho_t$ and $2L + 2\Delta + \rho_t$.
 Moreover, there exists a number $G\in \{0, L+\Delta, L-\Delta\}$ such that $|y-A.\hat{x}|$ are all $\leq \phi_t$ over these $(1-2\epsilon)~d$ measurements if we change $\hat{x_j}$ to $G$.
 \end{enumerate}
  \item [4)] set $\hat{x_{j}} \leftarrow G $, and go to 2 for next iteration.
  \item [5)] pick the set of $k$ significant elements of the candidate signal $\hat{x_T}$. Let $A'_{O(k \log n)\times k}$ be the sketch matrix $A$ restricted to those elements, output the solution of the optimization problem : \textit{find a vector $u$ to minimize} $|A'u-y|_2$.
	\end{algorithmic}
\end{algorithm}
In order to prove the algorithm we need the following definitions which are the generalization of the similar definitions in the exactly $k$-sparse case.
\begin{definition}
At each iteration t, we define $S_t$ an indicator of the difference between the estimate $\hat{x}$ and $x$  : $$S_t=\{j|\hat{x_j}\mbox{ and } x_j \mbox{in different levels or large with different signs.}\}.$$ 
\end{definition}

\begin{definition}[gap]  At each iteration $t$, let $G_t$ be the
  set of measurement elements in which at least one 'significant' elements from $\hat{x}$ contributes : $$G_t=\{i| |y_i-\sum_{j=1}^n A_{ij}\hat{x_j}|_{\infty} > \lambda D \}.$$	
\end{definition}

\begin{theorem}[Validity of the algorithm \ref{ouralgrob}]
The first part of the algorithm will find the position and sign of the $k$ significant elements of the signal $x$ (or more discussion see \cite{XH1}).

\begin{proof}
This is very similar to the proof of the validity of the exactly $k$-sparse recovery algorithm. We will exploit the following facts.

\begin{itemize}
\item $x$ is almost $k-sparse$ so it has at most $k$ significant elements.  Initially $S_1=k$ and $G_1\leq kd$.
\item Since at each iteration only one element $x_j$ is selected, at each iteration $t$ there are at most $t$ elements $x_j$ such that both $x_j$ and $\hat{x_j}$ are in the significant level  with the same sign.
\item If $S_t < 2k$ then $S_{t+1} < 2k $ (Preservation Lemma), and by the  neighborhood theorem at each round $(1-2\epsilon)|S_t| d \leq G_t $.

\item If $S_t <2k$ by the neighborhood theorem there exists a node $\hat{x_j}\in S_t$ which is the unique 
node in $S_t$ that is connected to at least $(1-2\epsilon)d$ parity check nodes. This node is in $S_t$. It differs from its actual value in the significance level or at sign. In the first case the part a) of the recovering algorithm will detect and fix it and in the second case the part b) of the algorithm will detect and fix it. For further discussion please refer to \cite{XH1}.
\item As a direct result : $|G_{t+1}|\leq |G_t| - (1-4\epsilon)d$ . So 
after $T=\frac{kd}{(1-4\epsilon) d}$ iterations we will have $G_T=0$. Consequently 
$S_T=0$  after at most $2k$ iterations.
\end{itemize}
This means that after at most $2k$ iterations the set $S_T=\{j|\hat{x_j}\mbox{ and } x_j \mbox{in different levels or with different signs.}\} $ will be empty and hence the position of the $k$ largest elements in $\hat{x_T}$ will be the position of the $k$ largest elements in $x$.
\end{proof}
\end{theorem}
\textbf{Remark} This algorithm like the exact $k$-sparse counterpart needs at most $2k$ iterations. Now by exploiting the simple structure of the adjacency matrix of the expander graph, (again in a similar manner to the exact $k$-sparse case), since $G_1 \leq kd$ initially we only need to construct a binary search tree for $kd$ elements. Moreover, at each iteration $G_{t+1} < G_t$. So even though at each iteration $(1-2\epsilon) d$ nodes are deleted from the tree and at most $2\epsilon d$ possibly new nodes are added to the tree the size of the tree never increases. Hence each iteration can be implemented efficiently in $O(d \log d)$ time complexity, and the algorithm will find the position of the $k$ largest elements of $x$ in $O(k \log n \log \log n)$ with a very small overhead. Note that the right-regularity assumption was only to make the analysis simpler and is not necessary.

Knowing the position of the $k$ largest elements of $x$ it will be easier to recover a good $k$-sparse approximation. Based on the $RIP-1$ property of the expander graph we propose a way to recover a good approximation for $x$ in a time sub-linear in $n$. We need the following lemma which is a direct result of the $RIP-1$ property of the expander graphs and is proved in \cite{berinde}
\begin{lemma}
\label{RIPR}
Consider any $u \in  R^n$ such that $|Au|_1 =b$, and let $S$ be any set of $k$ coordinates of $u$. Then we have:
\begin{equation}
 |u_S|_1 \leq \frac{b}{d(1-2\epsilon)}+\frac{2\epsilon}{1-2\epsilon}|u|_1.
\IEEEnonumber
\end{equation}and:
\begin{equation}
 \frac{1-4\epsilon}{1-2\epsilon}|u_S|_1 \leq \frac{b}{d(1-2\epsilon)}+\frac{2\epsilon}{1-2\epsilon}|u_{\bar{S}}|_1.
\IEEEnonumber
\end{equation}
\end{lemma}
Using Lemma \ref{RIPR} we will prove that the following minimization will recover a $k$-sparse signal very close to the original signal:
\begin{theorem}[Final recovery] Suppose $x$ is an almost $k$-sparse signal and $y=Ax$ is given where $y\in R^m$ and $m=O(k \log n)$. Also suppose $S$ is the set of the $k$ largest elements of $x$. Now let $A'$ be a submatrix of A containing columns from the positions of the $k$ largest elements of $x$, so $A'$ is a $O(k \log n) \times k$ dimensional matrix. Hence the following minimization problem can be solved in $O(poly(k, \log n))$ time complexity and will recover a $k$-sparse signal $v$ with very close norm-1 distance to the original $x$:
\begin{equation}
\mbox{min } |A'v-y|_2
\IEEEnonumber
\end{equation}
\proof Suppose $v$ is the recovered signal. Since $v$ is $k$-sparse we have $Av=A'v$ and hence: 

\begin{IEEEeqnarray}{rcl}
	|Av-Ax|_1&=&|Av-y|_1 \IEEEnonumber \\ &=& |A'v-y|_1 \IEEEnonumber\\ & \leq & \sqrt{m} |A'v-y|_2 \IEEEnonumber \\ &\leq& \sqrt{m} |A'x_S-y|_2 \IEEEnonumber \\ &=& \sqrt{m}|Ax_S-Ax|_2 \IEEEnonumber \\ &\leq& \sqrt{m } \lambda D \sqrt{m} \IEEEnonumber \\ &\leq& \sqrt{m^2 } \lambda D \IEEEnonumber \\ &=& mD\lambda = nd \lambda.
\end{IEEEeqnarray} 

The first two equations are only definitions. The third one is the Cauchy-Schwartz inequality. The fourth one is from the definition of $v$ and the last one is due to the almost-$k$-sparsity of $x$. Now by setting $u=x-v$ in Lemma \ref{RIPR} and since $v$ is $k$-sparse and $x$ is almost $k$-sparse with the same positions, we will have:
\begin{IEEEeqnarray}{rcl}
\frac{1-4\epsilon}{1-2\epsilon}|u_S|_1 &\leq& \frac{|Ax-Av|_1}{d(1-2\epsilon)}+\frac{2\epsilon}{1-2\epsilon}|u_{\bar{S}}|_1
\IEEEnonumber \\ &\leq& \frac{n \lambda}{(1-2\epsilon)}+\frac{2\epsilon}{1-2\epsilon}|u_{\bar{S}}|_1 \IEEEnonumber \\ &\leq& \frac{n\lambda}{(1-2\epsilon)}+\frac{2\epsilon}{1-2\epsilon}n\lambda
\IEEEnonumber \\ &=&O(n \lambda). \IEEEnonumber
\end{IEEEeqnarray}
As a result, since the signal is almost $k$-sparse, the value of $n \lambda$ will be negligible and hence  the recovered $k$-sparse signal and the $k$ largest elements of the original signal will be very close to each other. So the result will be a $k$-sparse signal approximating the original almost $k$-sparse signal with very high precision.
\end{theorem}


\textbf{Remark:} Recall that the right-regularity assumption is just for making the analysis simpler. As we mentioned before, it is not necessary for the first part of the algorithm. For the second part, it is used in the inequality $|Av-Ax| \leq \sqrt{m} |Ax_S-Ax|_2$.

However, denoting the i-th row of $A$ by $A_i$, we have 
$$|Ax_S-Ax|_2 = \sqrt{m} \sqrt{ \sum_{i=1}^m (A_i(x_S-x))^2 }
                     \leq \sqrt{m} \sqrt{ \sum_{i=1}^m (\lambda D_i)^2 }$$

where $D_i$ denotes the number of ones in the i-th row of $A$. (In the right regular case, $D_i = D$, for all i.)

Therefore:

$$|Ax_S-Ax|_2 \leq \sqrt{m} \lambda \sum_{i=1}^mD_i
                     = \sqrt{m} \lambda nd$$

The only difference with the constant $D_i$ case is the extra $\sqrt{m}$. But this does not affect the end result.

\section{Conclusion}
\label{sec:conc}
In this paper we used a combinatorial structure called an expander graph,
in order to perform deterministic efficient compressed sensing and
recovery. We showed how using
expander graphs one needs only $O(k \log n)$ measurements and the
recovery needs only $O(k)$ iterations. Also we showed how the
expansion property of the expander graphs, guarantees the full
recovery of the original signal. Since random graphs are with high
probability expander graphs and it is very easy to generate random
graphs, in many cases we might use random graphs instead. However, we
showed that in cases that recovery guarantees are needed, just with
a little penalty on the number of measurements and without affecting
the number of iterations needed for recovery, one can use another
family of expander graphs for which explicit constructions exists. We
also compared our result with a very recent result by Indyk et al
\cite{newindyk}, and showed that our algorithm has advantages in terms
of the number of required sketch measurements, the recovering complexity, and the simplicity of the algorithm in terms of the practical implementation. Finally, we showed how the algorithm can be modified to be robust. In order to do this we slightly modified the algorithm by using right-regular high quality expander graphs to find the position of the $k$ largest elements of an almost $k$-sparse signal. Then exploiting the robustness of the $RIP-1$ property of the expander graphs we showed how this information can be combined with efficient optimization methods to find a $k$-sparse approximation for $x$ very efficiently. However, in the almost $k$-sparsity model that we used non-sparse components should have "almost equal" magnitudes. This is because of the assumption that $L>k\Delta$ which restricts the degree of deviation for significant components. As a result, one important future work will be finding a  robust algorithm based on more general assumptions. Table
\ref{table} compares our results with the previous papers.

\begin{table}[!t]
\caption{Different Expander based recovery algorithms}
\center
\label{table}
\begin{tabular}{|c|c|c|c|c|c|}
\hline
Paper & R/D & Explicit & Sketch(m)  & \# Iterations & FullRecovery\\
\hline
\cite{CRT} & R & No & $O(k \log n)$ & LP ($O(n^3)$) & Yes: RIP-2\\
\cite{berinde} & D & No & $k \log\left(\frac{n}{k}\right)$ & LP ($O(n^3)$) & Yes:RIP-1\\
\cite{XH1}  & D & Yes & $\Theta(n)$ & $O(k)$ & Yes:RIP-1 \\
\cite{XH2} & D & No & $O(k\log n)$ & $O(k \log n)$ & Yes: RIP-1 \\
Theorem:\ref{recovery} &D & No & $O\left(k \log\left(\frac{n}{k}\right)\right)$ & $O(k)$ & Yes:RIP-1\\
Theorem:\ref{explicit} &D & Yes & $O(k^2 (\log n)^2)$ & $O(k)$ & Yes:RIP-1 \\
\hline
\end{tabular}
\end{table}

\end{document}